\documentclass[letter]{aa}     % for the letters
\usepackage{graphicx}
\usepackage{txfonts}
\usepackage{lipsum}
\usepackage{subcaption}         % necessary for continued figures, example in section 3
\usepackage{lscape}             % to rotate a single page table, example in appendix.
\usepackage{placeins}           % useful with \FloatBarrier, to keep 
\usepackage{url} %this package should fix any errors with URLs in refs.

\makeatletter
\let\linenumbers\relax
\let\endlinenumbers\relax
\makeatother

\begin{document}

   \title{Comparing simulated and observed particle energy distributions through magnetic reconnection in Earth's magnetotail}

   \author{N. Reisinger\inst{1} \and F. Bacchini \inst{1,2}}

   \institute{Centre for mathematical Plasma Astrophysics, Department of Mathematics, KU Leuven, Leuven, Belgium
   \and Royal Belgian Institute for Space Aeronomy, Solar-Terrestrial Centre of Excellence, Uccle, Belgium}

  \abstract
   {Magnetic reconnection is an explosive process that accelerates particles to high energies in Earth's magnetosphere, offering a unique natural laboratory to study this phenomenon. }
   {This study investigates how well data-driven fully kinetic simulations can reproduce the ion and electron energy distributions observed during a reconnection event by the Magnetospheric Multiscale (MMS) mission.} 
   {We performed fully kinetic 2D simulations initialized with plasma parameters derived from the MMS event and compared the resulting ion and electron energy distributions with observations. Key numerical and physical parameters were systematically varied to assess their influence on the resulting particle spectra.}
   {The simulations capture the overall shape and evolution of nonthermal energy distributions for both species, but generally underestimate the very high-energy tail of the electron spectrum. Variations in numerical parameters have negligible effects on the resulting spectra, while the initial upstream temperatures instead play a more pronounced role in reproducing the observed distributions.}
   {We present a novel analysis of data-driven fully kinetic simulations of MR, showing that key aspects of particle acceleration can be captured, while also highlighting the limitations of 2D simulations and the need for more realistic (e.g.,\ 3D) setups to  reproduce the observed particle energization accurately.}


   \titlerunning{Comparing Simulated and Observed Particle Distributions through Magnetic Reconnection}
   \authorrunning{Reisinger \& Bacchini}
    
   \maketitle

%%%%%%%%%%%%%%%%%%%%%%%%%%%%%%%%%%%%%%%%%%%%%%%%%%%%%%%%%%%%%%
\section{Introduction}

Magnetic reconnection (MR) is a ubiquitous plasma process occurring in many space and astrophysical environments. During MR, the topology of the local magnetic field is rapidly rearranged, and magnetic energy is converted to kinetic particle energy, potentially leading to significant particle acceleration \citep[e.g.,][]{Yamada_2010}. The dominant mechanisms for particle acceleration during MR are Fermi, betatron, and direct acceleration by parallel electric fields \citep[e.g.,][]{Northrop_1963, Birn_2012, Dahlin_2020, Li_2021}, while the relative contribution of each mechanism depends on the ambient plasma conditions \citep{Oka_2023}. An important feature of particle acceleration often associated with MR is the formation of a nonthermal power-law tail in the energy distribution $ f \propto E^{-p}$, with $E$ the particle energy and $p$ a spectral index determined by the local plasma conditions. Nonthermal particle spectra associated with MR were observed in numerical simulations with broadly varying physical parameters, ranging from trans- and ultrarelativistic conditions in astrophysical scenarios \citep[e.g.,][]{Sironi_2014, Guo_2014} to nonrelativistic regimes such as solar flares \citep[e.g.,][]{Li_2019_power_law, Zhang_2021}.

In comparison with more distant astrophysical objects, the day- and nightside of Earth's magnetosphere offers a unique opportunity for in situ measurements of MR and the associated particle dynamics. Due to the relatively simple magnetic-field configuration in Earth’s magnetotail, numerous reconnection events observed in that region have been used to study the associated particle acceleration \citep[e.g.,][]{Oieroset_2002, Zhou_2018, Ergun_2018, Ergun_2020, Zhong_2020, Cohen_2021, Oka_2022, Rajhans_2025}. Especially with the launch of NASA's Magnetospheric Multiscale (MMS) mission \citep{Burch_2016}, measurements of plasma parameters at high spatial and temporal resolutions became available. In contrast, relatively few fully kinetic simulations (i.e.,\ including electron physics) representing Earth's magnetotail have been presented in the literature \citep[e.g.,][]{Lapenta_2020}, with most studies instead focusing on reconnection under conditions typical of solar flares \citep[e.g.,][]{Li_2015,Li_2019_mass_ratio, Li_2019_power_law}. Notably, \citet{Nakamura_2018} performed a simulation in which the necessary input parameters to set up the initial state were extracted from an observed reconnection event in Earth's magnetotail. In this study, virtual spacecraft probes were used to compare field data, temperatures, and densities with in situ satellite measurements, showing remarkable agreement. However, no comparison was made between the simulation and observational data on particle energetics and acceleration. In general, only a few studies have compared energy distributions between simulations and observations; an example is \citet{Zhou_2018}, who compared electrons with THEMIS observations. However, no study has yet performed a quantitative comparison of the energy distribution of both particle species between MMS observations and fully kinetic simulations, capturing electron and ion physics and their feedback on the electromagnetic dynamics.
In this study, we carry out a number of fully kinetic simulations initialized with data from an observed magnetotail reconnection event and focus on evaluating how our runs can reproduce the measured energy distribution of electrons and ions. We perform a parameter study to explore how variations in plasma temperatures and the simulation setup affect the resulting particle energization.

%%%%%%%%%%%%%%%%%%%%%%%%%%%%%%%%%%%%%%%%%%%%%%%%%%%%%%%%%%%%%%

\section{Simulation setup}
\label{subsec:simset}

We conducted fully kinetic 2D simulations with the particle-in-cell code ECsim/RelSIM \citep{Lapenta_2017, Bacchini_2023}, which in the nonrelativistic limit conserves energy exactly (to machine precision). The simulations were initialized using a double Harris current sheet \citep{Harris_1962}, periodic boundary conditions along both axes, and a small initial perturbation to expedite the onset of MR. All simulation setups discussed below are based on the MR event observed on July 11, 2017, which was previously studied with fully kinetic simulations by \citet{Nakamura_2018}. In that work, a notable agreement was found between fluid plasma quantities measured by virtual spacecraft probes and the MMS3 observation. In the same spirit, the input parameters used in our runs are taken from observational data, described in Appendix~\ref{sec:Observations}. 

Our reference simulation, run $R0$ (see Appendices~\ref{sec:normalization}, \ref{sec:simres}, and \ref{sec:Table}), is based on the input parameters from \citet{Nakamura_2018} without major modifications, except for the omission of the guide field, whose strength is 3$\%$ of the upstream field and is therefore neglected. For the upstream magnetic-field strength ($B_0$), density ($n_0$) and ion and electron temperatures ($T_{0,i}$ and $T_{0,e}$), the following values were used: $B_0 = 12$~nT, $n_0 = 0.3$~cm$^{-3}$, $T_{0,i} = 1500$~eV, and $T_{0,e} = 500$~eV. The upstream plasma was initialized with a uniform temperature and density. Furthermore, the initial density ($n_d$) and ion and electron temperatures ($T_{d,i}$ and $T_{d,e}$) in the current sheet (providing initial pressure equilibrium) are given by $n_d/n_0 = 3$, $T_{d,i}/T_{0,i} = 3$, and $T_{d,e}/T_{0,e} = 3$. The simulation box spans $L_x\times L_y=23\times69 (d_{0,i})^2$ with a grid spacing $\Delta x=\Delta y =0.12 d_{0,i}$, where $d_{0,i}$ is the ion skin depth calculated with $n_0$, and a time step, $\Delta t$, of 0.03$\omega_{0,i}^{-1}$, where $\omega_{0,i}$ is the ion plasma frequency based on $d_{0,i}$. $\Delta x$ and $\Delta y$ were chosen such that, at the initial state, each cell resolves at least one electron skin depth, which is $d_{0,e}=0.125 d_{0,i}$ for run $R0$ and similarly applied in all other runs. The initial half-thickness of the current sheet is $\delta = 0.35d_{0, i}$, the ion-to-electron mass ratio is $m_i/m_e$ = 64, and in the whole simulation domain 2.8 $\times 10^7$ particles are used, which are initialized with a Maxwellian particle distribution. Details of the simulation initialization are given in Appendix~\ref{sec:normalization}, and in Appendix~\ref{sec:simres}, Figure~\ref{fig:E_ref} shows the energy evolution and particle energy distributions for simulation $R0$. 

All other simulations ($R1$--$R7$) deviate from $R0$ in at least one parameter; these variations are highlighted in bold in Table~\ref{tab:sim_overview} in Appendix~\ref{sec:Table}. To investigate the influence of the mass ratio $m_i/m_e$ on particle energization, we set up runs $R1$ and $R2$, with $m_i/m_e =25$ and 100, respectively. Due to the change in $m_i/m_e$, modifications were made for the grid spacing, $\Delta x$, and the time step, $\Delta t$, to ensure that we resolved one inertial length of electrons with one cell in the initial simulation state and to keep the ratio $\Delta t / \Delta x$ between 0.2 and 0.25, as in $R0$. For runs $R3$, $R4$, and $R5$, the box size of the simulation domain is varied: in $R3$ and $R4$, $L_x$ or $L_y$ are increased by 25$\%$ while keeping the length of the other box side constant, while in $R5$ both $L_x$ and $L_y$ are increased by 25\%. %, while all other parameters in these three runs are the same as $R0$. 
For the last two runs, $R6$ and $R7$, the temperatures of the background plasma were modified. Specifically, in $R6$ the ion temperature, $T_{0,i}$, is kept constant and the electron temperature, $T_{0,e}$, is decreased to 150~eV to obtain $T_{0,i}/T_{0,e} = 10$, instead of 3 as in $R0$. Hence, the total plasma-$\beta$, $\beta = \beta_i + \beta_e = n_0 k  8 \pi (T_{0,i} + T_{0,e})/B_0^2$, was reduced from 0.16 to 0.13 due to the change in $T_{0,e}$. The conditions of the upstream plasma in the magnetotail are not exactly known, as MMS3 crosses the EDR along the MR exhausts \citep{Torbert_2018}. Hence, the temperatures ($T_{0,i}$, $T_{0,e}$) were estimated with a different approach in $R7$: in this run, ion and electron temperatures were obtained via a Maxwellian fit of the core of the observed distributions, as we discuss in later sections. This leads to $T_{0,i} = 5100$~eV and $T_{0,e} = 1000$~eV, as well as to $T_{0,i}/T_{0,e} = 5.1$, and agrees with the median temperature values within the time window: 996~eV for the electron temperature and 5122~eV for the ion temperature. Furthermore, in this run, the magnetic-field strength, $B_0$, is increased to 21~nT to maintain the same plasma-$\beta$ as in $R0$.

%%%%%%%%%%%%%%%%%%%%%%%%%%%%%%%%%%%%%%%%%%%%%%%%%%%%%%%%%%%%%%
\section{Results}
\label{sec:results}

\subsection{Parameter study}

\begin{figure*}[ht]
    \centering
    \includegraphics[width =0.8\textwidth]{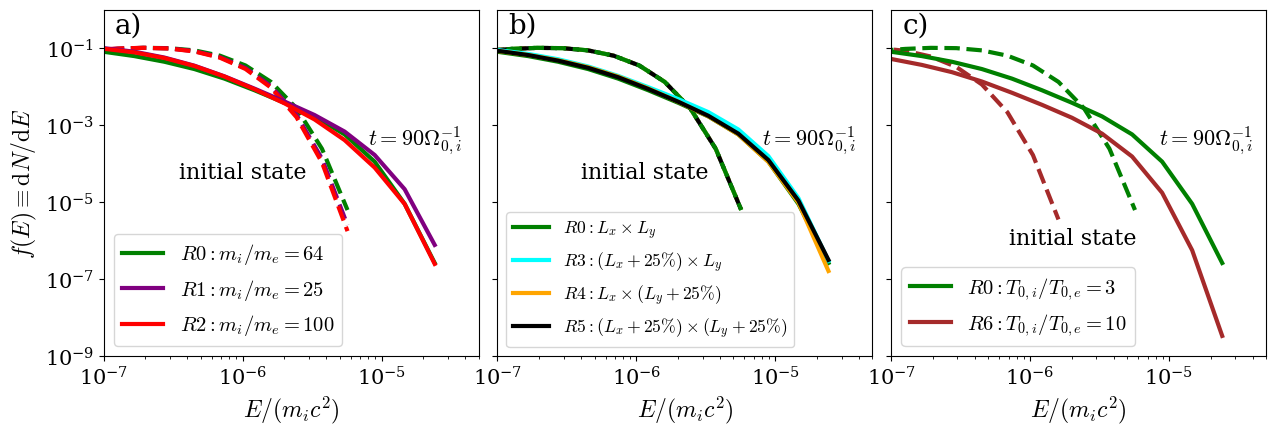}
    \caption{Electron energy distributions for different simulations, considering the variation in specific parameters: the mass ratio, $m_i/m_e$, in panel a); the simulation domain size, $L_x \times L_y$, in panel b); and the initial temperature ratio, $T_{0,i}/T_{0,e}$, in panel c). The dashed lines show the initial energy distributions and the solid lines represent the distributions at $t = 90\Omega_{0,i}^{-1}$.}
    \label{fig:E_parameters}
\end{figure*}

Fully kinetic simulations impose artificial limitations compared to real processes, such as a finite box size or the usage of a reduced ion-to-electron mass ratio. In order to study the influence of these limitations on particle energization during MR, certain parameters have been changed in simulations $R1$--$R7$ and compared with the reference simulation, $R0$: the mass ratio, $m_i/m_e$, the size of the simulation domain, $L_x \times L_y$, and the initial electron temperature, $T_{0,e}$ (see Table~\ref{tab:sim_overview}). The electron energy distributions for different simulations are shown in Figure~\ref{fig:E_parameters}, where the distributions for the initial state ($t = 0$, dashed lines) and at a later stage ($t = 90\Omega_{0,i}^{-1}$, solid lines) are shown, where $\Omega_{0,i}$ is the ion cyclotron frequency calculated with $B_0$. The normalization of the energy distributions is discussed in Section~\ref{subsec:comparison}. In Figure~\ref{fig:E_parameters}a), the electron energy distributions for different $m_i/m_e$ are shown. In runs $R1$ and $R2$, $m_i/m_e$ was varied from 64 (as used in $R0$) to 25 and 100, respectively. 
In R1 and R2 we observe a weak tendency to reach larger electron energies for smaller $m_i/m_e$, showing that the resulting energy spectra are essentially insensitive of the mass ratio. In Figure~\ref{fig:E_parameters}b), the electron energy distributions for different simulation domain sizes are shown. $R0$ is the reference box with $L_x \times L_y = 23 \times 69 (d_{0,i})^2$, while in $R3$ and $R4$, $L_x$ or $L_y$ are increased by 25$\%$, respectively. In $R5$, the box size is increased in both directions by 25$\%$, leading to $L_x \times Ly = 29 \times 86 (d_{0,i})^2$. The energy distributions for both species show essentially no differences. In our last parameter scan, the initial temperature ratio, $T_{0,i}/T_{0,e}$, was increased from 3 to 10 (run $R6$), while $T_{0,i}$ was kept at the same initial value as in $R0$. The ion energy distribution shows no difference between the two cases and is thus not shown here. Figure~\ref{fig:E_parameters}c) instead shows the resulting electron energy distribution of the reference simulation, $R0$, compared with the result of run $R6$. In this case, the initial electron energy distribution of the two simulations differs. The initial electron temperature in simulation $R6$ is lower compared to simulation $R0$. However, at $t = 90 \Omega_{0,i}^{-1}$, the energy distributions in both simulations only exhibit a small difference. Notably, the higher $T_{0,i}/T_{0,e}$ (run $R6$) leads to stronger energization: while electrons in run $R6$ start at significantly lower energies than in $R0$, their final state reaches energies comparable to those attained in $R0$.

In summary, the numerical parameters $m_i/m_e$ and $L_x \times L_y$ have no strong effect on the resulting energy spectra, supporting the validity of the parameters used in $R0$. In contrast, the upstream particle temperatures must be selected with greater caution due to their significant influence on the particle energization. Accordingly, a different method was used to estimate the upstream temperatures in run $R7$ to quantitatively reproduce the observations, as is described below.

\subsection{Comparison between observations and simulations}
\label{subsec:comparison}

\begin{figure}[h]
    \centering
    \includegraphics[width=0.98\columnwidth]{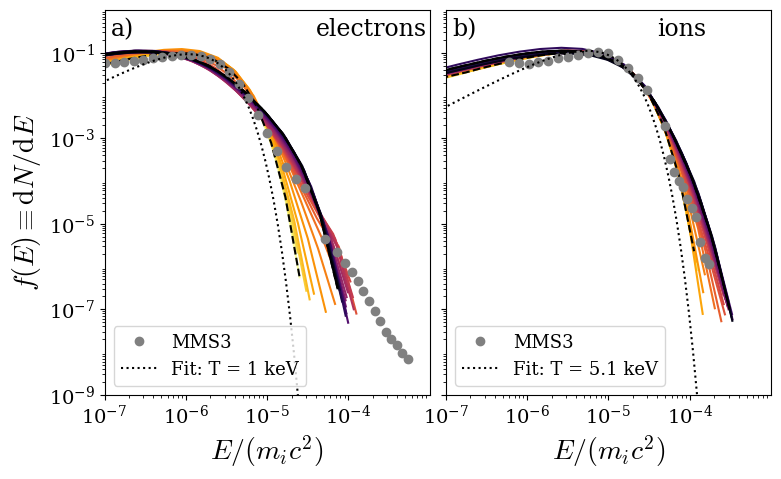}
    \caption{Energy distributions from MMS3 observations (gray dots) and simulation R7. Solid lines show the simulated electrons (a) and ions (b) at different times until $135\Omega_{0,i}^{-1}$, with later times in darker shades. Dotted lines represent Maxwellian fits of the core of the observed distributions, and dashed lines show the simulated energy distributions at initialization.}
    \label{fig:E_comparison}
\end{figure}

The ultimate goal of this study is to compare particle energization resulting from MR observed at the Earth's magnetotail with the outcome from fully kinetic simulations, and especially reproducing the high-energy electron population. One challenging aspect is ensuring consistent units and scaling of physical quantities between observations and simulations. Hence, all the energies, $E$, from observations and simulations, are normalized by $m_ic^2$. However, the comparison of the energy distributions from the two techniques posed additional difficulties, as the units of the differential directional number flux from MMS measurements are challenging to replicate in simulations. Thus, particle fluxes within a certain energy range are normalized by the total number of particles across all energy ranges in both observations and simulations. With this normalization, observations and simulations can be compared quantitatively, as is shown in Figure~\ref{fig:E_comparison}, where the solid lines are from run $R7$ and the circles from observations. Notably, the energy distribution functions in Figures~\ref{fig:E_ref} and \ref{fig:E_parameters} use the same scaling. In Figure~\ref{fig:E_comparison}, we show the energy distribution functions from run R7 until $t=135\Omega_{0,i}^{-1}$. 
The initial temperature values for run $R7$ in Figure~\ref{fig:E_comparison} are based on Maxwellian fits of the observed distributions, represented by the dotted lines. While the Maxwellian fits are in good agreement with the thermal core measured in the observations, matching the nonthermal part of the observed spectrum requires reconnection to push particles to higher energies. Over the evolution of the simulation, the initially Maxwellian populations develop a nonthermal, high-energy part in their distributions. The observed energy range is reached at approximately $50\Omega_{0,i}^{-1}$, and from that point the distributions slowly converge to a final, converged state. Considering the distributions throughout the simulation, electrons reach higher energies relative to their initial state and exhibit a broader energy distribution than ions. At $t = 50\Omega_{0,i}^{-1}$, the high-energy electron population ($E/(m_ic^2) \gtrsim 10^{-5}$) contains $5\%$ of their total population and contributes to $24\%$ of the total kinetic electron energy, while $6\%$ of the ions are in the high-energy population ($E/(m_ic^2) \gtrsim 4 \times 10^{-5}$) and represent $21\%$ of the total kinetic ion energy. Overall, our simulations show the development of nonthermal populations in the energy distribution for electrons and ions, attaining energies broadly consistent with those observed in the Earth's magnetotail, except for a marked discrepancy in the high-energy electron tail. The latter is discussed below. 

%%%%%%%%%%%%%%%%%%%%%%%%%%%%%%%%%%%%%%%%%%%%%%%%%%%%%%%%%%%%%%

\section{Discussion and conclusions}
\label{sec:disc}

We performed eight fully kinetic 2D simulations of MR in Earth's magnetotail, based on satellite data measured during a reconnection event by MMS. To initialize our reconnection runs, the upstream plasma parameters were directly inferred from observational data, and key numerical parameters of the simulation setup, including the mass ratio, box size, and initial temperature ratio, were systematically varied to assess their effect on the system evolution. For the first time, the particle energization obtained from data-driven fully kinetic simulations was compared with MMS observations of both ion and electron energy spectra. Our simulations quantitatively retrieve the observed energies and trend for both ions and electrons, except for the very high-energy range of the electrons, which remains underrepresented.

In our parameter scan, we investigated the effect of the ion-to-electron temperature ratio. While low plasma-$\beta$ simulations have been performed in previous studies to investigate MR, the focus was primarily on scenarios with equal electron and ion temperatures -- conditions that are more appropriate for investigating MR in solar flares \citep[e.g.,][]{Li_2019_mass_ratio, Li_2019_power_law, Zhang_2024}, but not representative of Earth's magnetosphere. Our results show a significant influence of the initial particle temperatures on the energization (see Figure~\ref{fig:E_comparison}c)); therefore, this distinction is significant. In terms of the mass ratio, \citet{Li_2019_mass_ratio} found a minimal effect on the resulting energy distributions under solar flare conditions, consistent with our findings. 

Our reference simulation ($R0$) is based on the initialization from \citet{Nakamura_2018}; that work demonstrated good quantitative agreement in fluid quantities between a virtual spacecraft probe and MMS measurements. Here, we focused on particle energization, which, however, for $R0$ did not reproduce the observed energy distributions from MMS3 well. A much better agreement was obtained by estimating the upstream temperatures more accurately, as we did in our final simulation ($R7$). Inferring the upstream plasma conditions from observations is challenging, as the MMS spacecraft crossed the electron diffusion region along the outflow direction, limiting information about the background plasma. Therefore, we initialized the temperature values with a different approach and used a Maxwellian approximation to fit the thermal core of the observed distributions, as is shown in Figure~\ref{fig:E_comparison}. The simulated ion and electron energies in $R7$ are consistent with observations, broadly retrieving the correct ranges for the nonthermal populations. However, we note that the actual upstream temperatures may be lower than our estimates, since MMS likely encountered plasma already influenced by reconnection. Carefully calibrating the upstream conditions away from the reconnection region using more observational data will be the subject of future work.

In our results, the very high-energy end of the observed electron spectrum, where only a few particles reside, is not accurately captured. This could be attributed to a number of factors, including the numerical parameters used in our simulations. Furthermore, it is generally accepted that in nonrelativistic reconnection scenarios with a low (or zero) guide field, Fermi-type acceleration dominates particle energization \citep{Dahlin_2017, Li_2018, Li_2019_power_law}. This process is driven by curvature drift associated with the relaxation of magnetic tension. In 2D simulations, particles can get trapped in long-lived magnetic islands, because of the limited movement of particles across magnetic field lines \citep[e.g.,][]{Jokipii_1993, Jones_1998}, which prohibits additional magnetic-field destruction and particle energy gain until primary islands merge with each other \citep{Li_2015, Li_2017}. In more realistic 3D scenarios, particles can initially become confined within flux ropes extending in the out-of-plane direction, until flux-rope kink instabilities are triggered. Flux-rope kinking can destroy these structures and create a more turbulent state, allowing particles to escape and experience additional Fermi acceleration, potentially reaching higher energies \citep{Zhang_2021}. 
To attain more realistic results and capture the very high-energy end of the electron spectra, in future studies we aim to enhance our simulations by fine-tuning the numerical parameters, conducting fully 3D simulations to overcome 2D limitations, and considering the conditions for the onset of kink instabilities. Another limitation with respect to realistic magnetotail conditions is the use of periodic rather than open boundary conditions. However, since the goal of this study is to quantify how a finite amount of magnetic energy is converted to particle kinetic energy within a controlled, closed system, periodic boundaries are sufficient to capture the local energization processes. In the future, we shall also conduct careful comparisons between open and closed boundary simulations to assess the influence of periodicity on electron energization. Finally, in this study, particle distributions were measured globally over the whole domain, while spacecraft data represent localized measurements in space. While our approach is sound for investigating the high-energy range of the distributions, globally dominated by the highly energetic particles in the current sheet, a better comparison would require local measurements. Therefore, future work will include the use of virtual spacecraft probes to enable a more realistic comparison of particle energization between simulations and MMS measurements. 

%%%%%%%%%%%%%%%%%%%%%%%%%%%%%%%%%%%%%%%%%%%%%%%%%%%%%%%%%%%%%%

\section*{Data availability}
All the observational data are publicly available from the MMS Science Data Center. 
Simulation data are generated with ECsim/RelSIM, as described in Section~\ref{subsec:simset}, and available at \citet{Reisinger_2025}. 
%%%%%%%%%%%%%%%%%%%%%%%%%%%%%%%%%%%%%%%%%%%%%%%%%%%%%%%%%%%%%%

\begin{acknowledgements}
      N.R.\ was supported by FWO Flanders fellowship, with ID no. 1106626N and \emph{ERC Advanced Grant TerraVirtualE} of Giovanni Lapenta (grant agreement No.\ 1101095310). Views and opinions expressed are however those of the authors only and do not necessarily reflect those of the European Union or the European Commission. Neither the European Union nor the European Commission can be held responsible for them.
    F.B.\ acknowledges support from the FED-tWIN programme (profile Prf-2020-004, project ``ENERGY'') issued by BELSPO, and from the FWO Junior Research Project G020224N granted by the Research Foundation -- Flanders (FWO). 
    The resources and services used in this work were provided in part by the VSC (Flemish Supercomputer Center), funded by the Research Foundation - Flanders (FWO) and the Flemish Government. We acknowledge EuroCC Belgium for awarding this project access to the LUMI supercomputer, owned by the EuroHPC Joint Undertaking, hosted by CSC (Finland) and the LUMI consortium.
\end{acknowledgements}

%%%%%%%%%%%%%%%%%%%%%%%%%%%%%%%%%%%%%%%%%%%%%%%%%%%%%%%%%%%%%%

\bibliographystyle{aa.bst} % style aa.bst
\bibliography{bib.bib} % your references Yourfile.bib

%%%%%%%%%%%%%%%%%%%%%%%%%%%%%%%%%%%%%%%%%%%%%%%%%%%%%%%%%%%%%%%
\begin{appendix}
%%%%%%%%%%%%%%%%%%%%%%%%%%%%%%%%%%%%%%%%%%%%%%%%%%%%%%%%%%%%%%%
\onecolumn

\section{Observational data}
\label{sec:Observations}
The used observations are from a well-reported reconnection event observed by NASA's Magnetospheric Multiscale (MMS) mission \citep{Burch_2016}. MMS was launched to study MR on kinetic scales at the Earth's magnetosphere within two distinct regions: at the dayside magnetopause and in the nightside magnetotail. During the second phase, when the MMS mission was investigating the Earth's magnetotail, the MMS spacecraft crossed an electron diffusion region (EDR) on July 11, 2017 at 22:34:03 during a reconnection event \citep[e.g.,][]{Genestreti_2018, Nakamura_2018, Torbert_2018, Nakamura_2019}. Our study uses publicly available level-2, burst-mode data from MMS3 during 22:33 - 22:35 UT. The particle data comes from the Fast Particle Detector (FPI; \citet{Pollock_2016}) and the Energetic Particle Detector (EPD), specifically from the Fly's Eye Energetic Particle Sensor (FEEPS; \citet{Blake_2016}) and Energetic Ion Spectrometer (EIS;\citet{Mauk_2016} ). The temperature data and low-energy ranges for both species are from FPI, while EIS and FEEPS cover the high-energy ranges for ions and electrons, respectively. The energy data used in Section~\ref{subsec:comparison} is averaged during the time window for each energy level, and similar to previous studies \citep[e.g.,][]{Oka_2022}, only energies higher than 100~eV and 500~eV for electrons and ions, respectively, were considered. This removes data that might be impacted by artificial effects and ensures consistent energy ranges in observations and simulations.

\section{Simulation initialization}    
\label{sec:normalization}
To initialize the simulations using observational constraints, all physical parameters are converted to code units. In the employed code, the speed of light $c$, the ion charge $q_i$, and the ion mass $m_i$ are set to 1. The initial background number density is set to $n_0 = (4\pi)^{-1}$. The background magnetic field strength is specified via the Alfv\'en speed, $v_{A,j} = B_0/{\sqrt{4\pi n_0 m_j}}$ for species $j$, where $n_0$ and $B_0$ correspond to the observed background density and magnetic field strength defined in Section~\ref{subsec:simset}. Assuming $m_i \gg m_e$ and using the above normalization, the Alfv\'en speed is roughly equal to the ion Alfv\'en speed $v_{A,i} = B_0$ in code units. The thermal velocities are defined as $v_{th,j} = \sqrt{k T_j/ m_j}$, for each species $j$. The drift velocities of the current sheet species are estimated from Amp\`ere's law and the pressure balance condition.

\section{Overview of simulation results}
\label{sec:simres}

\begin{figure}[!hb]
    \centering
    \includegraphics[width=0.65\linewidth]{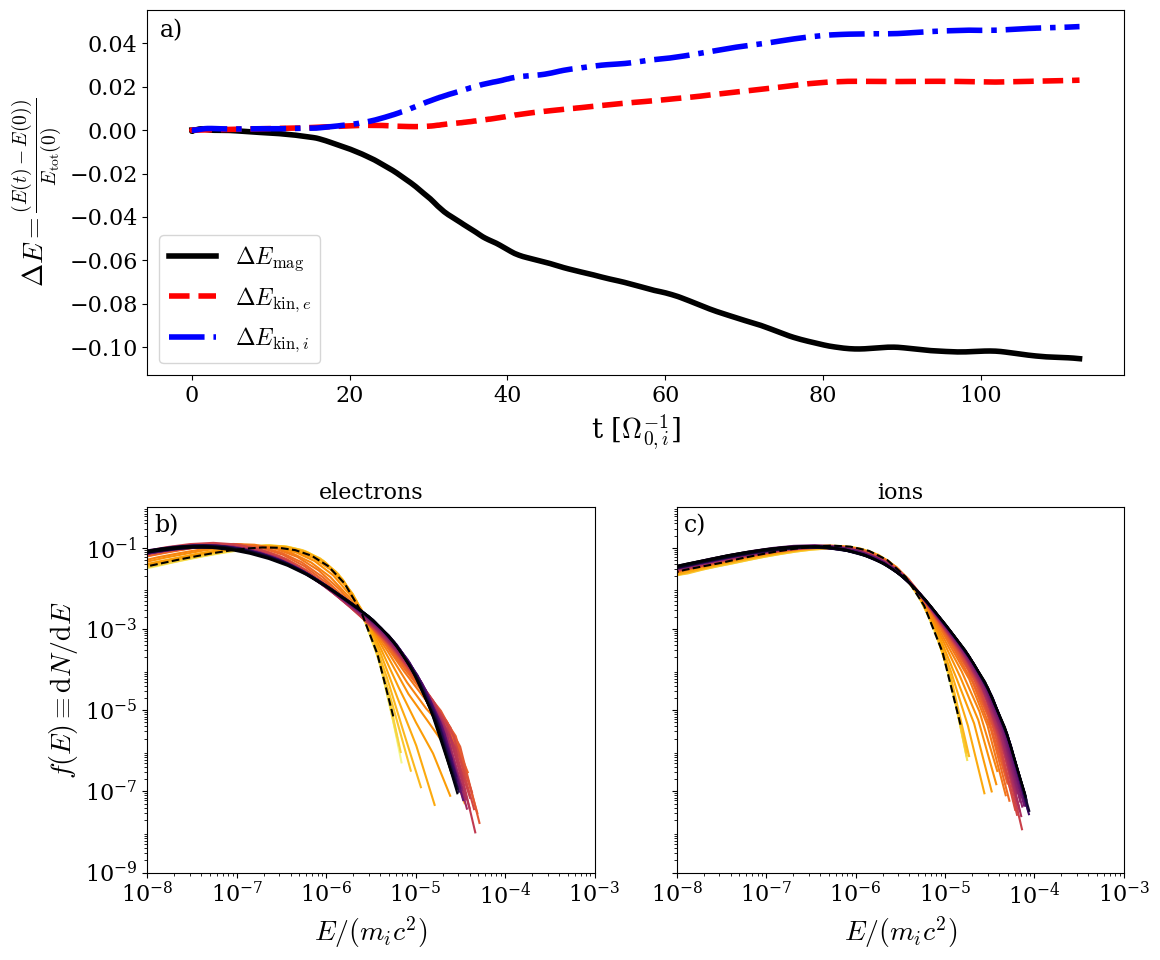}
    \caption{Evolution of different energy quantities throughout the simulation $R0$ in a), calculated as the energy change $\Delta E = (E (t) - E(0))/E_\mathrm{tot}(0)$, with $E$ the specific energy quantity (magnetic or kinetic) and $E_\mathrm{tot}$ the total energy within the system. Energy distributions $f(E)\equiv\mathrm{d}N/\mathrm{d}E$ are shown for electrons (b)) and ions (c)) at different times throughout the same simulation. Both panels show the evolution until $t = 105 \Omega_{0,i}^{-1}$, with darker shades indicating later times. The dashed lines show the initial distribution of each particle species.}
    \label{fig:E_ref}
\end{figure}

The normalized time evolution of various energy components throughout simulation $R0$ is shown in Figure~\ref{fig:E_ref}a), and exhibits the typical temporal dynamics of energy transfer during reconnection processes. After around $15\Omega_{0,i}^{-1}$, the total magnetic energy $E_\mathrm{mag}$  starts to decrease, marking the onset of the nonlinear reconnection phase. This is followed by a gradual increase in the ion kinetic energy $E_{\mathrm{kin},i}$ and subsequently in the electron kinetic energy $E_{\mathrm{kin},e}$ --- indicating the conversion of magnetic energy to kinetic particle energy. At around $80\Omega_{0,i}^{-1}$, the rate of depletion of magnetic energy slows significantly, similar to the enhancement in both kinetic energies. %The change in total energy $\Delta E_\mathrm{tot}$ remains constant throughout the simulation, with only minor deviations that are negligible compared to the changes observed in the other energy components. Notably, this is a unique characteristic of the used code, see Subsection \ref{subsec:simset}. 
Figure~\ref{fig:E_ref}b) and c) show the evolution of the energy distribution for ions and electrons, respectively, from the same simulation, $R0$. Both species deviate substantially from the initial distribution (dashed lines) over time, as indicated by the darker solid lines representing later simulation times. During the run, both species are energized to higher energies and a nonthermal population appears. All other simulations conducted show the same qualitative behavior in the energy evolution and distributions and are therefore not shown here.

\section{All simulations discussed}
\label{sec:Table}

Table \ref{tab:sim_overview} summarizes all simulations in this study and clarifies which parameters were modified with respect to the reference run (R0).

\begin{table}[ht!]
    \caption{Overview of all eight simulations, with deviations from the reference simulation ($R0$) in bold.}
    \centering
    %\begin{tabular}{c|c|c|c|c|c|c|c|c|c|c}
    \begin{tabular}{c c c c c c c c c c c}
        Run & $m_i/m_e$ & $\Delta x/d_{0,i}$  & $\Delta t/\omega_{0,i}^{-1}$& $L_x/d_{0,i}$ & $L_y/d_{0,i}$ & $\beta$ & $B_0$~$[$nT$]$ & $T_{0,i}$~$[$eV$]$ & $T_{0,e}$~$[$eV$]$ & $T_{0,i}/T_{0,e}$\\
        \hline
         $R0$ & 64 & 0.12 & 0.03 & 23 & 69 & 0.16 & 12 & 1500 & 500 & 3.0 \\
         \hline
         $R1$ & $\boldsymbol{25}$ & $\boldsymbol{0.15}$ & 0.03 & 23 & 69 & 0.16 & 12 & 1500 & 500 & 3.0 \\
         $R2$ & $\boldsymbol{100}$ & $\boldsymbol{0.08}$ & $\boldsymbol{0.02}$ &23 & 69 & 0.16 & 12 & 1500 & 500 & 3.0 \\
         \hline
         $R3$ & 64 & 0.12 & 0.03 & $\boldsymbol{29}$ & 69 & 0.16 & 12 & 1500 & 500 & 3.0 \\
         $R4$ & 64 & 0.12 & 0.03 & 23 & $\boldsymbol{86}$ & 0.16 & 12 & 1500 & 500 & 3.0 \\
         $R5$ & 64 & 0.12 & 0.03 & $\boldsymbol{29}$ & $\boldsymbol{86}$ & 0.16 & 12 & 1500 & 500 & 3.0 \\
         \hline
         $R6$ & 64 & 0.12 & 0.03 & 23 & 69 & $\boldsymbol{0.13}$ & 12 & 1500 & $\boldsymbol{150}$ &  $\boldsymbol{10.0}$ \\
         $R7$ & 64 & 0.12 & 0.03 & 23 & 69 &  0.16 & $\boldsymbol{21}$ & $\boldsymbol{5100}$ & $\boldsymbol{1000}$ &  $\boldsymbol{5.1}$ \\
    \end{tabular}
    \tablefoot{Table includes the mass ratio between ions and electrons $m_i/m_e$, grid spacing $\Delta x=\Delta y$, time step $\Delta t$, domain size $L_x \times L_y$, total plasma-$\beta$, background magnetic field $B_0$, and electron and ion temperatures $T_{0,e}$ and $T_{0,i}$. Length scales are normalized to the ion inertial length $d_{0,i}$ calculated with the background density $n_0$, and time scales are normalized to the inverse ion plasma frequency $\omega_{0,i}^{-1}$.}
    \label{tab:sim_overview}
\end{table}

\end{appendix}
\end{document}